\documentclass[aps,prl,reprint,showpacs,twocolumn]{revtex4-1}
\usepackage{graphicx}
\usepackage{textcomp}
\usepackage{subfigure}
\newcommand{\ket}[1]{\left|#1\right\rangle}
\newcommand{\SState}{\text{S}_{1/2}}
\newcommand{\PState}{\text{P}_{3/2}}
\newcommand{\DState}{\text{D}_{5/2}}

\begin{document}

\title{Doubly-heralded single-photon absorption by a single atom}

\author{J. Brito$^{1}$}
\author{S. Kucera$^{1}$}
\author{P. Eich$^{1}$}
\author{P. M\"uller$^{1}$}
\author{J. Eschner$^{1}$}
\email{juergen.eschner@physik.uni-saarland.de}

\affiliation{
$^1$Universit\"at des Saarlandes, Experimentalphysik, Campus E2 6, 66123 Saarbr\"ucken, Germany\\
}

\date{\today}

\begin{abstract}

We report on a single-photon-to-single-atom interface, where a single photon generated by Spontaneous Parametric Down Conversion (SPDC) is absorbed by a single trapped ion. The photon is heralded by its time-correlated partner generated in the SPDC process, while the absorption event is heralded by a single photon emitted in its course. Coincidence detection marks doubly-heralded absorption, enabling photon-to-atom quantum state transfer \cite{Kurz2014, Mueller2013}. Background in the coincidence measurement is strongly suppressed by a new method that discriminates real absorption events from dark count-induced coincidences. 
\end{abstract}


\maketitle

A major goal in quantum technologies is to integrate quantum communication and information processing into quantum networks \cite{Kimble2008}. One promising approach is using single atoms and photons as nodes and channels of the network \cite{Ritter2012, Boozer2007, Schug2013}. Single photons are able to transmit quantum information between different nodes of the network, while single atoms serve as high-fidelity, long-storage-time memories; single trapped ions, in particular, may additionally be used as quantum processors \cite{Langer2005, Stute2012, Schindler2013, Hucul2015}. In this context, the use of heralding photons for repeaters and memories schemes is very promising \cite{Moehring2007, Duan2012}: successful entanglement transfer or information storage onto a single atom is signaled by the detection of a photonic herald emitted as a consequence of the process. The detection of such heralds permits one to discriminate successful events and thereby enables high-fidelity protocols, e.g. for photon-to-atom quantum state transfer \cite{Weber2009, Kurz2014} or distant ion entanglement \cite{Moehring2007}. 

Previously we have demonstrated the absorption of single SPDC photons in a single ion, signaled by the onset of fluorescence, i.e., by a quantum jump \cite{Piro2011}: absorption of a photon excites the ion out of a metastable state into a fluorescence cycle, and a large number of emitted photons heralds the absorption event with near $100\%$ efficiency. This method has also been used to herald single-photon interaction between two distant ions \cite{Schug2013} and to manifest the polarization entanglement of a SPDC photon pair in single-photon absorption \cite{Huwer2013}. By generating fluorescence, however, the state of the ion after the single-photon absorption is not preserved, and transferred quantum information is lost. This limitation is overcome when a single scattered photon is used to herald the absorption process \cite{Lloyd2001, Weber2009, Mueller2013, Sangouard2013}. In \cite{Kurz2014} we implemented a specific protocol based on such heralding that enabled high-fidelity transfer of a photon polarization onto an atomic qubit, and that may also be used for photon-to-atom entanglement mapping \cite{Sangouard2013}. While this experimental demonstration was carried out with laser photons, here we extend its implementation to the absorption of single photons from an SPDC source, thus proving the feasibility of photon-to-atom quantum state conversion with our interface. 

The principle of our single-photon-to-single-atom interface is as follows: a single trapped ion is exposed to a single resonant photon, generated via SPDC and heralded by its partner; successful absorption is heralded by the detection of a single Raman-scattered photon released as a result of the absorption. Coincident detection of the partner SPDC photon and the photon emitted by the ion marks doubly-heralded single-photon absorption by a single atom. Furthermore, we present a method for discriminating real absorption events from dark-count induced coincidences, tackling and eliminating the main source of background.

\subsection{Experimental setup}

\begin{figure}[htbp]
\centering
\includegraphics[width=0.47\textwidth]{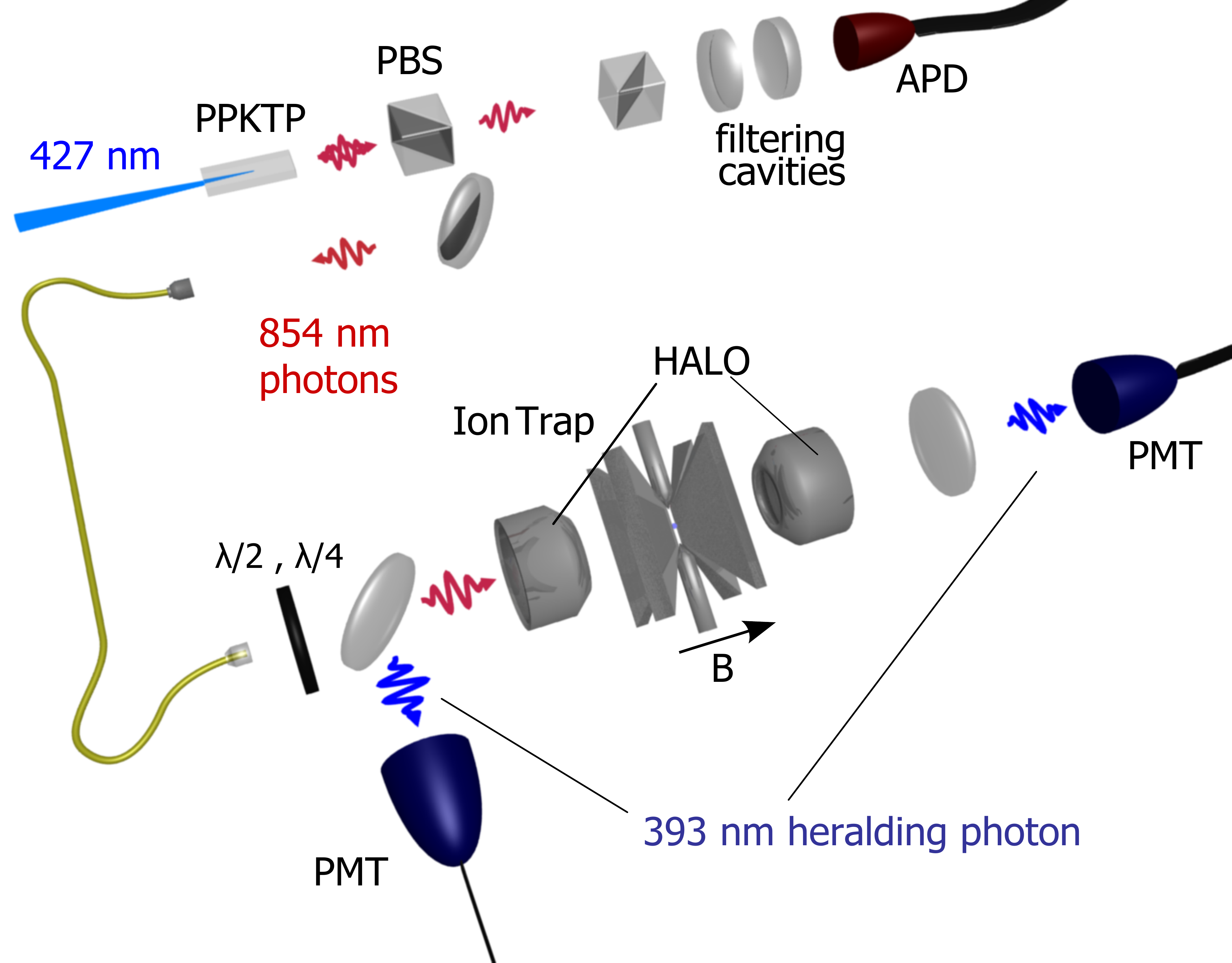}
\caption{Experimental setup. The top part shows the SPDC source of resonant, narrowband, heralded single photons; the lower part shows the ion trap setup consisting of a linear Paul trap with two high numerical aperture laser objectives (HALOs). A single-mode optical fiber connects the two setups. Further details are given in the text.}
\label{fig:setup}
\end{figure}

The experimental apparatus, sketched in Fig.~\ref{fig:setup}, consists of the SPDC source setup \cite{Haase2009, Piro2009} and the ion trap setup and is similar to the one used in \cite{Piro2011} and \cite{Huwer2013}. Degenerate photon pairs at 854~nm are created collinearly by spontaneous parametric down-conversion in a PPKTP (periodically poled potassium titanyl phosphate) non-linear crystal. The crystal is pumped with a continuous laser beam at 427~nm obtained by second harmonic generation of an amplified 854~nm diode laser (Toptica, TA-SHG pro), which is stabilized to the ${\rm D}_{5/2} \to {\rm P}_{3/2}$ transition in the $^{40}$Ca$^+$ ion via a transfer lock \cite{Rohde2010a}. The two photons of a pair have perpendicular polarizations due to type II phase-matching. Their polarization entanglement is projected out when they are split by a polarizing beam splitter (PBS), in order to use them as a heralded single-photon source. The vertically polarized photons are coupled into a single-mode fiber and transmitted to the ion trap setup for absorption; their polarization at the ion is adjusted using waveplates ($\lambda/4$, $\lambda/2$). Utilizing the time-correlation of the SPDC pair, the horizontally polarized photons are used to signal the presence of their partners. They are filtered by two cascaded Fabry-Perot cavities, which only transmit the $22$ MHz bandwidth of the ${\rm D}_{5/2} \to {\rm P}_{3/2}$ transition, and which are actively stabilized to resonance. Transmitted photons are detected with an avalanche photo diode (APD, Perkin-Elmer, $\sim30\%$ efficiency). This selects only ion-resonant photons from the broadband ($\sim 200$ GHz) SPDC emission. More details are explained in Refs.~\cite{Haase2009, Piro2009}. 

\begin{figure}[htbp]
\centering
\includegraphics[width=0.3\textwidth]{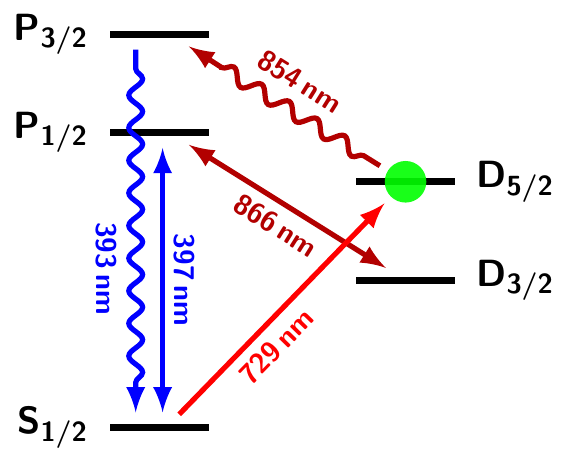}
\caption{Relevant levels and transitions of the $^{40}\text{Ca}^+$ ion. Lasers at 397~nm and 866~nm are used for cooling and fluorescence detection. Lasers at 729~nm and 854~nm are used for optical pumping and state preparation. The single ion is prepared in the D$_{5/2}$ metastable state for absorption of 854~nm SPDC photons. Absorption heralds are emitted at 393~nm.}
\label{fig:levels} 
\end{figure}

The linear Paul-type ion trap holds a single $^{40}$Ca$^+$ ion, whose relevant levels and transitions are shown in 
Fig.~\ref{fig:levels}. The ion is initially prepared in the ${\rm D}_{5/2}$ level, from where it can be excited to ${\rm P}_{3/2}$ by absorption of a single 854~nm photon. Following absorption, it decays to the ${\rm S}_{1/2}$ ground state with 93.5\% probability (according to the branching ratio of the ${\rm P}_{3/2}$ state), thereby releasing a single photon at 393~nm. If detected, this photon heralds the single-photon absorption. For efficient ion-photon coupling, two in-vacuum high numerical aperture laser objectives (HALOs, $NA=0.4$) are placed along one of the radial trap axes, which coincides with the quantization axis (magnetic field direction, $\vec{B}$ in Fig.~\ref{fig:setup}). One of the HALOs is used to focus the 854~nm SPDC photons onto the ion, while both of them collect the 393~nm absorption heralds, which are extracted with dichroic mirrors and detected by two photomultiplier tubes (PMTs, Hamamatsu). Including collection and detection efficiency, $1.86\%$ of the emitted heralds are finally detected. The ion is cooled, pumped and prepared for absorption by several lasers as described in Fig.~\ref{fig:levels}. 

\subsection{Experimental sequence}

The experimental sequence runs at 401 Hz repetition rate and consists of several phases. First, Doppler cooling is applied for 100~$\mu$s using 397~nm and 866~nm laser light, which brings the ion to the Lamb-Dicke limit. After optical pumping to the $|m=-\frac{1}{2}\rangle$ Zeeman substate of the S$_{1/2}$ ground state for 60~$\mu$s, the ion is prepared for the absorption of 854~nm SPDC photons by coherent excitation to the $|m=-\frac{5}{2}\rangle$ substate of the D$_{5/2}$ metastable manifold (lifetime $\tau_{\rm D}=1.17$~s) with a 7~$\mu$s long $\pi$-pulse of a narrow-band $729$~nm laser. Subsequently, a detection window is open for $2$~ms, while photons from the SPDC source are sent to the ion at a rate of $\sim 3500~{\rm s}^{-1}$ resonant photons. These photons are set to be $\sigma^+$ polarized in the ion's reference frame, thereby exciting the ion to the $\ket{\PState,m=-\frac{3}{2}}$ level, which corresponds to the transition with the highest absorption probability between D$_{5/2}$ and P$_{3/2}$. During the interaction phase, detection events of 854~nm photons on the APD and 393~nm photons on the PMTs are recorded and time-tagged for their later correlation analysis. After an experimental cycle of cooling, pumping, preparation, and interaction has finished, a series of 729~nm pi-pulses is used for discrimination of the background, as explained below. The effective single-photon-single-atom interaction time is 68\% of the measurement time, resulting from the 2~ms interaction period per 2.5~ms cycle repetition, and reduced by the periodically applied stabilization of the cavity filters~\cite{Haase2009, Piro2009}.

\subsection{Experimental results}

\begin{figure}[tbp]
\includegraphics[width=0.49\textwidth]{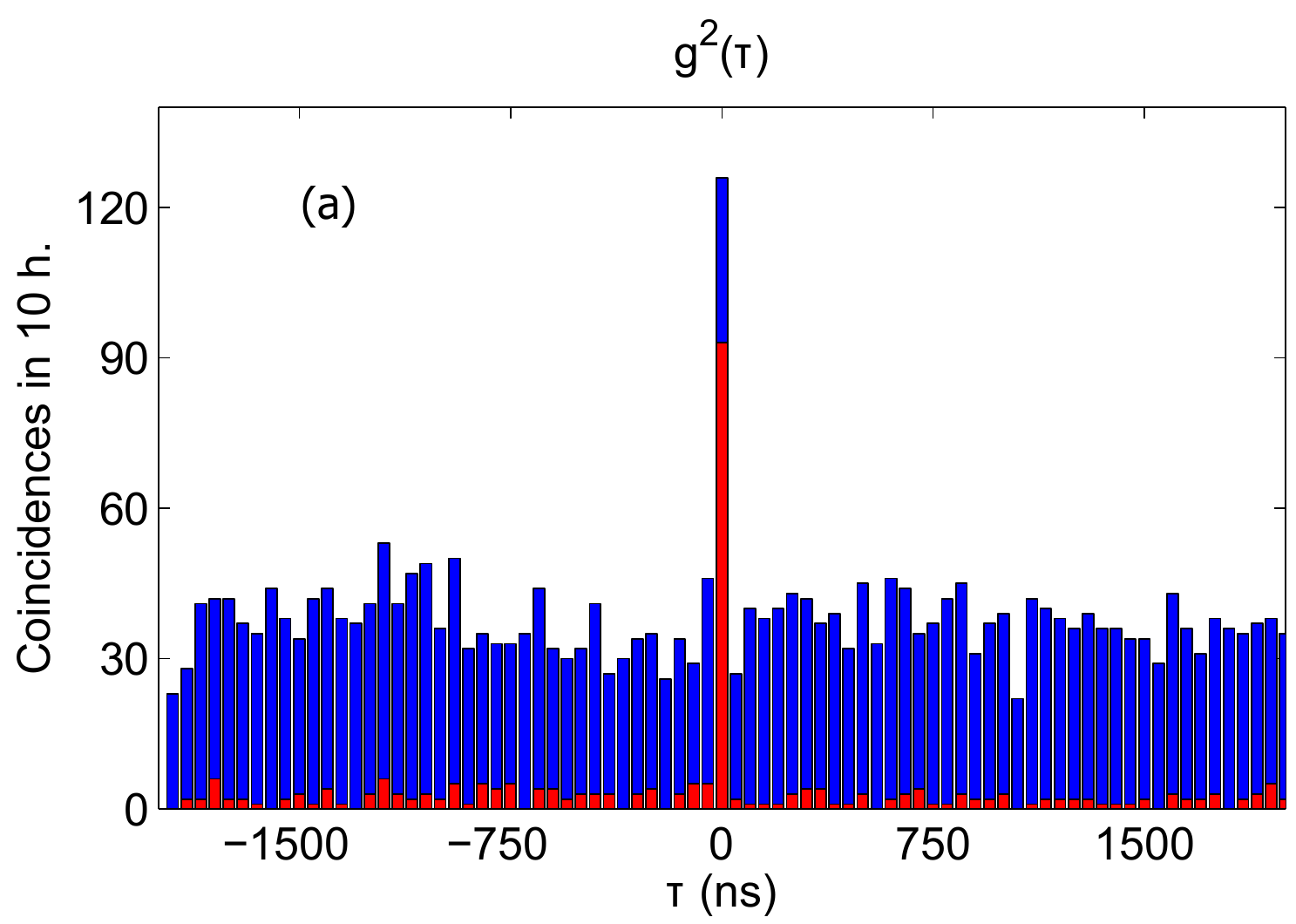}
\includegraphics[width=0.49\textwidth]{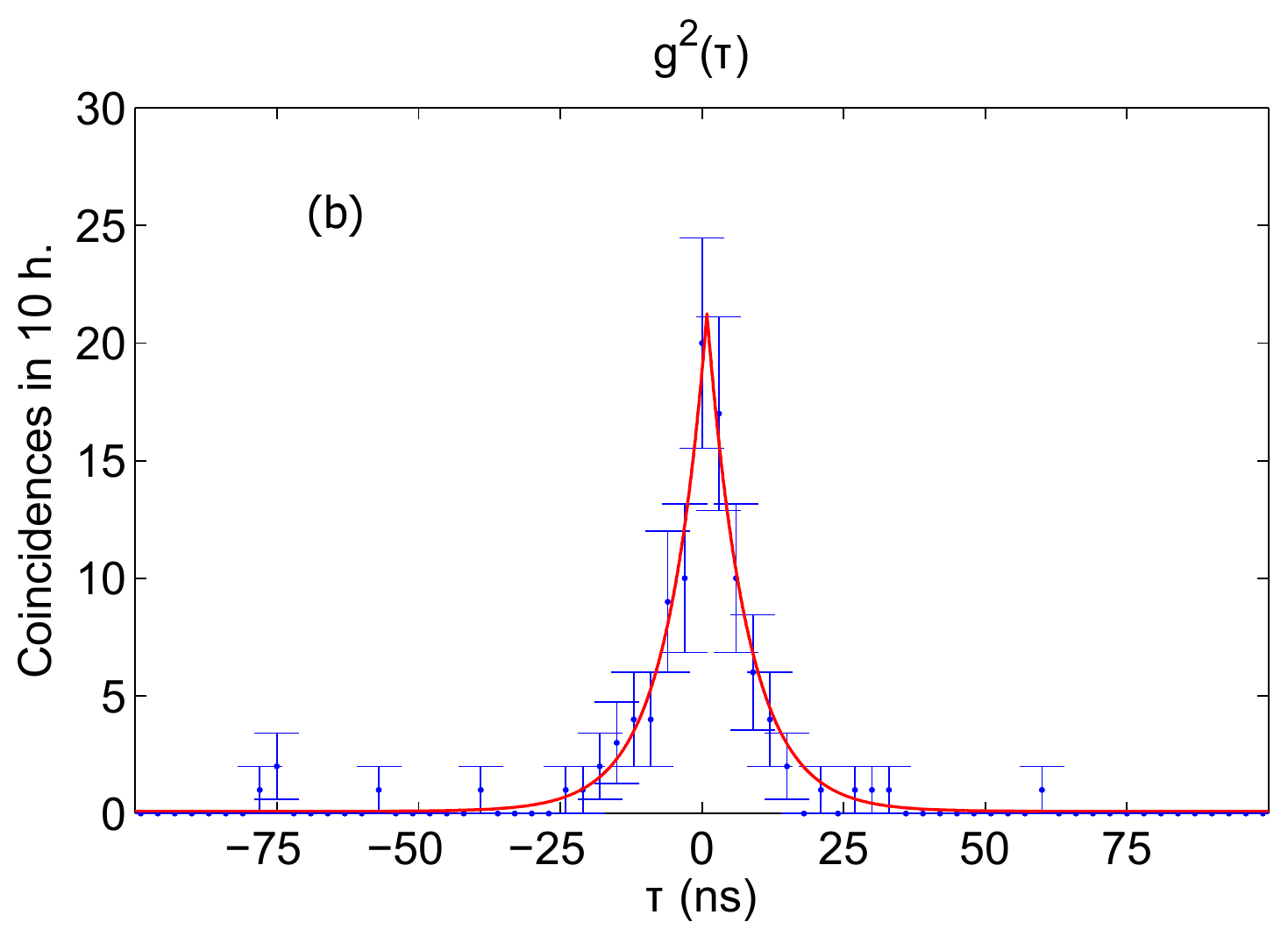}
\caption{(a) Time correlation between photon heralds (854~nm) and absorption heralds (393~nm) with 50~ns binning and 10~hrs total measurement time (24.540~s interaction time). The peak at zero time delay corresponds to doubly-heralded single-photon absorption. The blue and red histograms show, respectively, the same data without and with additional background reduction (see text). (b) Coincidence peak after background reduction and with 3 ns binning. The red line represents the expected correlation function.\label{fig:histo}}
\end{figure}

Time correlation histograms of the detections are shown in Fig.~\ref{fig:histo}. The peak around zero time delay corresponds to coincident photon detection on the APD and on one of the PMTs, and thereby marks doubly-heralded single-photon absorption. In the histogram of Fig.~\ref{fig:histo}(a) shown in blue, $89$ doubly-heralded absorption events in 10~hrs (24,540~s interaction time) were detected (background subtracted), meaning an average coincidence rate of $3.6 \cdot 10^{-3}$ per second. With an average background of $37.1$ counts per 50~ns bin and Poissonian noise, the signal-to-background ratio (SBR) amounts to 2.4 and the signal-to-noise ratio (SNR) is $14.6$. Higher time resolution of the coincidence peak using $3$ ns binning (Fig.~\ref{fig:histo}(b)) reveals a symmetric double-sided exponential of about 7~ns 1/e-time, originating from the temporal shapes of the two heralding photons: the shape of the 393~nm herald represents the lifetime of the P$_{3/2}$ state ($\tau_{\rm P}=7.2$~ns), while the shape of the 854~nm photons corresponds to the ring-down time of the filtering cavities, designed to match the atomic spectral shape \cite{Haase2009, Piro2009}. The red line in the plot represents the expected shape of the correlation function, showing very good agreement with the experimental findings. 

An absorption efficiency per source photon of  $\sim 1 \cdot 10^{-3}$ is derived for this experiment, which is higher than the value reported in \cite{Schug2013, Huwer2013}, which we attribute to better preparation of the ion in the $\ket{\DState,\text{m}=-\frac{5}{2}}$ state by using the 729~nm laser.

\subsection{Background reduction}

\begin{figure}[ht]
\includegraphics[height=0.45\columnwidth]{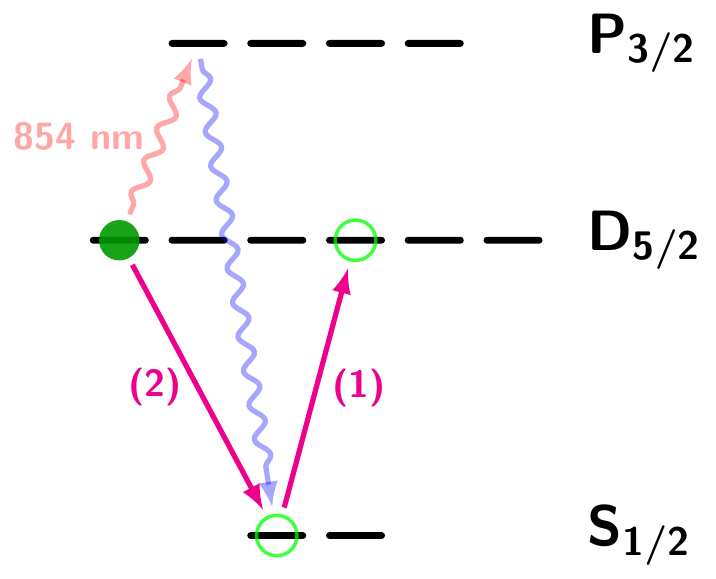}
\caption{Experimental scheme and background reduction. The scheme shows the absorption of an 854~nm photon (wavy red arrow) and subsequent emission of the photon at 393~nm that heralds the absorption process (wavy blue arrow). State measurement for background reduction employs two coherent pulses at 729~nm (arrows labeled (1) and (2)) after the single-photon interaction period, and subsequent state-selective fluorescence, as explained in the text. 
\label{fig:scheme}}
\end{figure}

In a further step of the data analysis, a discrimination method is included to eliminate the most important source of background, i.e., events where a photon detection on the APD is coincident with a dark count on the PMT. Reducing the background is important since it decreases the measurement time for a target SNR or SBR and increases the fidelity of quantum state transfer schemes \cite{Kurz2014}, by disregarding accidental coincidences that mimic a real absorption. Our method to achieve this is to measure the atomic state after the interaction and consider only those coincidence detections where the atom has indeed made the transition from $\ket{\DState,\text{m}=-\frac{5}{2}}$ to $\ket{\SState,\text{m}=-\frac{1}{2}}$. For this purpose, a series of laser pulses (altogether $\sim300~\mu$s) was added to the experimental sequence after the ion-photon interaction phase. First, as shown in Fig.~\ref{fig:scheme}, two 729~nm $\pi$-pulses are applied to the ion. The first pulse takes the population from $\ket{\SState,\text{m}=-\frac{1}{2}}$ to $\ket{\DState,\text{m}=+\frac{3}{2}}$, which is used as an auxiliary metastable state. This population corresponds to the case where absorption of an 854~nm photon has happened.  The following pulse transfers the population from $\ket{\DState,\text{m}=-\frac{5}{2}}$ to $\ket{\SState,\text{m}=-\frac{1}{2}}$, which corresponds to the case where no photon was absorbed. Subsequently, by switching on the cooling lasers for 250~$\mu$s, both cases are discriminated. The onset of fluorescence (ion turning bright) means that the ion remained in $\DState$ during the interaction, and no 854~nm photon was absorbed. In the case where an absorption event happened, the ion will stay dark. By selecting only those 393~nm heralds that go along with a dark ion, we disregard coincidence events triggered by dark counts of the PMTs. While the coherent pulses, which in effect swap the bright and the dark result of the state measurement, do not seem to be strictly necessary for the state discrimination, they avoid that spontaneous decay from D$_{5/2}$ to S$_{1/2}$ during the state measurement mimics an absorption event, and furthermore they are part of more extended method which will be explained below. 

The result for the background reduction method as described so far is shown in Fig.~\ref{fig:histo}(a), where the red bars correspond to the correlation measurement for the same data as the blue bars, but with post-selection of the relevant events. The average background per bin is reduced from $37.1$ to $2.3$ events, a $16$-fold improvement of the SBR, which translates into a $4.1$-fold increase of the SNR. After applying the background reduction as described, there are still two cases which account for most of the remaining background counts, namely spontaneous decay from D$_{5/2}$ to S$_{1/2}$, which happens with $0.17\%$ probability within the 2 ms of absorption window, and failed initial state preparation in D$_{5/2}$ ($\sim 1\%$). Since in these cases the ion population will be in S$_{1/2}$ after the interaction phase, they will be counted as successful absorptions events when they coincide with a double-herald detection. For the particular measurement of Fig.~\ref{fig:histo} these "false positives" are not critical, as verified by the low observed background (2.3 events for 50~ns bins, or 0.14 events for 3~ns bins, as in Fig.~\ref{fig:histo}(b)). Nevertheless, our background reduction method can be extended to also eliminate the background related to failed preparation: by applying a 397~nm laser pulse after the 729~nm preparation pulse, remaining S$_{1/2}$ population can be stored in the D$_{3/2}$ metastable state until the end of the interaction period. The coherent pulses after the interaction period leave the D$_{3/2}$ population untouched, such that it will correctly contribute to the bright cases, i.e. to the no-absorption result of the state measurement.

\subsection*{Conclusion}

As a summary, we have realized an interface where a single emitted photon heralds the absorption of a single SPDC photon by a single atom. This demonstrates that our protocol for high-fidelity heralded photon-to-atom quantum state transfer, presented in \cite{Kurz2014}, can be extended to single photons. Successful absorption is heralded by the coincident detection, within $<50$~ns, of the emitted photon at 393~nm and the SPDC partner photon at 854~nm. The time correlation displays the expected double-exponential temporal structure. Furthermore, we presented a way to efficiently reduce background counts by verifying the atomic state after the detection of a herald, which is important as the fidelity of photon-to-atom state conversion may be reduced by dark counts that mimic heralds. With small time overhead and no additional lasers, we decreased the background counts by a factor of $16$, leading to a $4.1$-fold improvement of the signal-to-noise ratio. These results add another proof-of-principle to constructing a quantum network toolbox with single photons and single atoms.

As an outlook, a brighter entangled photon source will allow us to implement full photon-to-atom quantum state transfer and to work towards entanglement mapping from photon pairs to distant atoms \cite{Mueller2013, Kurz2014, Sangouard2011}. Moreover, quantum-frequency conversion of heralding photons or non-degenerate SPDC pairs with one partner in the telecom range \cite{Lenhard2015} will connect atomic quantum memories and processors with long-haul quantum communication and with different quantum systems such as solid-state memories and color-center- or semiconductor-based single-photon sources \cite{Zaske2012}, in order to construct a multi-platform hybrid quantum network.

\begin{acknowledgments}
We acknowledge support by the BMBF (Verbundprojekt Q.com, CHIST-ERA project QScale), the German Scholars Organization / Alfried Krupp von Bohlen und Halbach-Stiftung, and the ESF (IOTA COST Action). J. Brito acknowledges support by CONICYT.
\end{acknowledgments}

\bibliography{refs}

\end{document}